\documentclass[twocolumn,nopacs,prl,amsmath,amssymb]{revtex4}
\usepackage{graphicx}% Include figure files
\usepackage{booktabs}
\usepackage{multirow}
\usepackage{amssymb}

\renewcommand{\vec}[1]{\mathrm{\mathbf{#1}}}

\begin{document}
\title{Surface Enhanced Raman Spectroscopy of Graphene}
\author{F. Schedin$^1$, E. Lidorikis$^2$,A. Lombardo$^3$,V. G. Kravets$^1$, A. K. Geim$^1$, A. N. Grigorenko$^1$, K. S. Novoselov$^1$}\email{kostya@manchester.ac.uk}\author{A. C. Ferrari$^3$}\email{acf26@eng.cam.ac.uk}
\affiliation{$^1$Department of Physics and Astronomy, Manchester University, Manchester, UK\\$^2$Department of Materials Science and Engineering, University of Ioannina, Ioannina, Greece\\$^3$Department of Engineering, University of Cambridge,Cambridge CB3 0FA, UK}

\begin{abstract}
Surface enhanced Raman scattering (SERS) exploits surface plasmons induced by the incident field in metallic nanostructures to significantly increase the Raman intensity. Graphene provides the ideal prototype two dimensional (2d) test material to investigate SERS. Its Raman spectrum is well known, graphene samples are entirely reproducible, height controllable down to the atomic scale, and can be made virtually defect-free. We report SERS from graphene, by depositing arrays of Au particles of well defined dimensions on graphene/SiO$_2$(300nm)/Si. We detect significant enhancements at 633nm. To elucidate the physics of SERS, we develop a quantitative analytical and numerical theory. The 2d nature of graphene allows for a closed-form description of the Raman enhancement. This scales with the nanoparticle cross section, the fourth power of the Mie enhancement, and is inversely proportional to the tenth power of the separation between graphene and the nanoparticle. One consequence is that metallic nanodisks are an ideal embodiment for SERS in 2d.
\end{abstract}
\maketitle
\section{Introduction}
Graphene is at the center of a significant research effort\cite{Nov306(2004),GeimRevNM, Nov438(2005),CastroNetoRev,charlier,Zhang438(2005)}. Near-ballistic transport at room temperature and high mobility\cite{Nov438(2005),Zhang438(2005),Nov315(2007),MorozovNov(2007),andrei,kimmob, kimmob2} make it interesting for nanoelectronics\cite{Han, Chen,Zhang86,Lemme}, especially for high frequency applications\cite{Yuming}. Furthermore, its transparency and mechanical properties are ideal for micro and nanomechanical systems, thin-film transistors, transparent and conductive electrodes, and photonics\cite{bunch,blake1,hernandez,eda,zhipcondmat,Hasan,gokus,krupke}. Graphene is also an ideal test-bed for some long-standing problems, such as the Raman spectra of carbon materials. Here we show that this conceptually simple material (due to its low-dimensionality) can be helpful in understanding the basics of Surface Enhanced Raman Scattering (SERS).

Graphene layers can be identified by inelastic\cite{ACFRaman} and elastic light scattering\cite{CasiraghiNL,BlakeAPL}. Raman spectroscopy allows monitoring of doping, defects, strain, disorder, chemical modifications and edges\cite{ACFRaman,CasiraghiAPL,Pisana,ACFRamanSSC,DasCM,Mohi,YanPrl2007,Leandro,PiscanecPRL,Cedge,Elias,Dasbila,Ferrari00,Graf2007,acftrans}. The Raman signal can be enhanced for flakes deposited on certain substrates, such as the common Si+SiO$_2$, due to interference in the SiO$_2$ layer, resulting into enhanced field amplitudes within graphene\cite{wang2008,son2009,gao2009,casiraghi2010}.

Another way to increase the Raman signal is to perform SERS experiments\cite{moskovits85,kneipp06}. SERS is widely used\cite{junsen,kneipp,fleischmann74}, and enhancements as large as 14 orders of magnitude can be achieved (enough for single-molecule detection\cite{nie97}). However, even though the technique is more than 30 years old\cite{fleischmann74}, the exact nature of SERS is still debated\cite{junsen}. Furthermore, the particular mechanism might be different, depending on whether the Raman processes involved are resonant or not. In principle, even a single metallic nanostructure, e.g., a metallic nanotip, can induce SERS at its apex, giving rise to the so-called tip-enhanced Raman scattering (TERS)\cite{novotny07,kawata09,cancado2009}. The most important feature that makes TERS so attractive is its capability of optical sensing with high spatial resolution beyond the light diffraction limits\cite{kawata07,achim}.

Most SERS-active systems studied to-date are based on random nanostructures, whose properties vary from experiment to experiment making quantitative comparison between theory and experiment difficult. Graphene offers a unique model system where SERS effects could be studied in detail. Its Raman spectrum is well known, being investigated in several hundreds papers in the past 4 years. Graphene samples are very reproducible and offer an atomic-precision control on the number of layers, thus allowing a smooth transition from a purely 2d case to a 3d one. Furthermore, as both resonant and non-resonant Raman scattering can be in principle possible, such as in chemically modified graphene\cite{Elias}, a distinction between different enhancement mechanisms could be made. Here we focus on the resonant case, where we believe the enhancement is mostly due to near-field plasmonic effects in the vicinity of metal particles\cite{cancado2009,moskovits85}.

\section{Experimental}
Graphene flakes are prepared on Si+300nm SiO$_2$ by micromechanical cleavage\cite{Nov306(2004)}. Single layer graphene (SLG) is identified by a combination of optical contrast\cite{CasiraghiNL,BlakeAPL} and Raman spectroscopy\cite{ACFRaman}. Electron beam lithography in combination with thin metallic film deposition (5nm Cr+ 80nm Au) and lift-off are utilized to prepare three sets of metallic dots, as well as a set of contacts for transport measurements, Fig 1. One set is placed directly on top of graphene, one partially covers graphene and partially rests on SiO$_2$, and the last completely on SiO$_2$. The dots sizes and the configurations of the arrays can be seen on Fig.1. During lift-off, metallic dots are slightly shifted from their lithographically defined positions, probably by capillary forces.  This indicates poor adhesion of Cr/Au dots onto graphene. Note that the dots on SiO$_2$ still occupy the positions defined by the lithography procedure.
\begin{figure}
\centerline{\includegraphics[width=70mm]{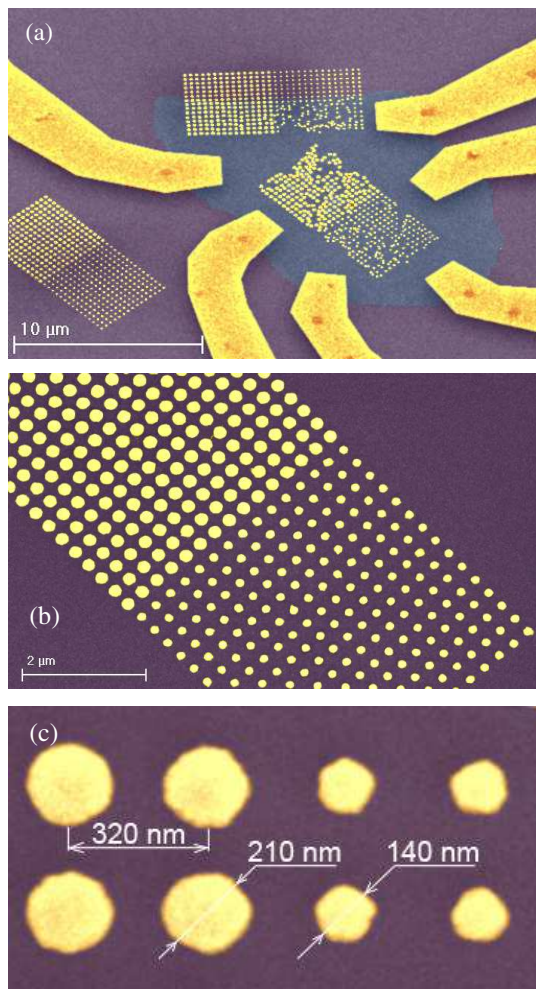}}
\caption{Scanning Electron Microscopy images (in false colours) of our SERS sample. False colors denote: purple - SiO$_2$; bluish - graphene; yellow - Au electrodes and dots. a) An overall image of the sample. b,c) Golden dots on SiO$_2$}
\label{fig:schematic1a}
\end{figure}

Raman spectra are recorded with a Renishaw RM1000 spectrometer, equipped with a piezoelectric stage (PI) able to shift the sample at nanometer steps. Line scans are recorded across the patterned arrays, as shown in Figs 2,3, for 488, 514 and 633nm excitation.
\section{Results and Discussion}
The Raman enhancement is defined as the ratio of the Raman intensity measured on the graphene covered by dots, compared to that measured outside the dots, but still on the graphene layer, Fig. 2.

Fig. 3 shows representative Raman spectra measured at 633nm. A clear enhancement is seen when comparing the patterned and the unpatterned graphene.
\begin{figure}
\centerline{\includegraphics[width=60mm]{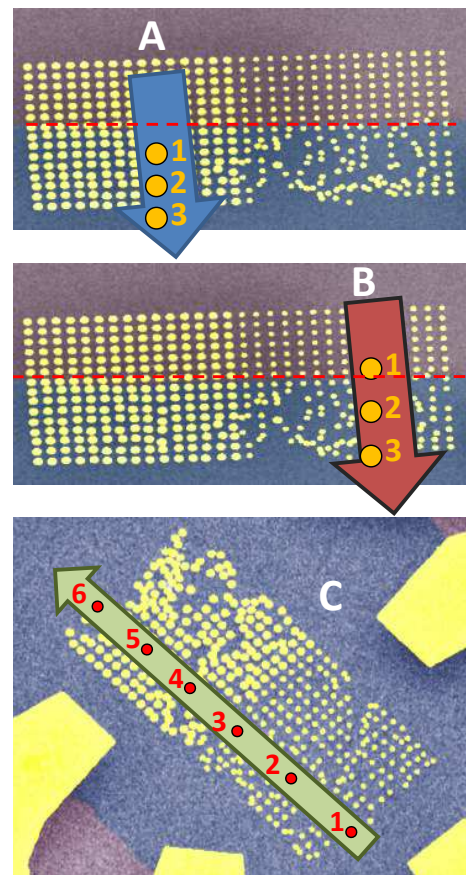}}
\caption{Raman linescans. Dotted red lines indicate the graphene edge. The intensity ratios are a) large dots: Point 1/Point 3; Point 2/Point 3; b) small dots: Point 1/Point 3; Point 2/Point 3; c) large dots: Point 4/ Point 6; Point 5/Point 6; small dots: Point 3/Point 1; Point 2/Point 1}
\label{fig:linescan}
\end{figure}
\begin{figure}
\centerline{\includegraphics[width=70mm]{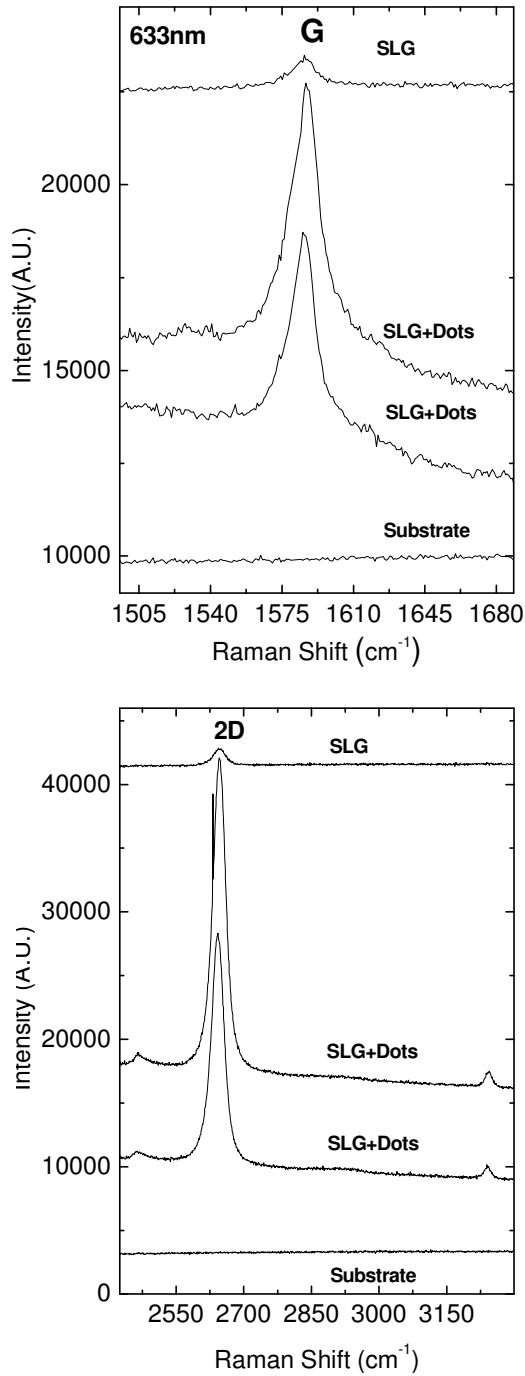}}
\caption{Representative Raman spectra measured across a line scan moving from outside to inside the Au dot patterned area for 633 excitation. A clear enhancement of all peaks is seen}
\label{fig:spectra}
\end{figure}
\begin{figure}
\centerline{\includegraphics[width=70mm]{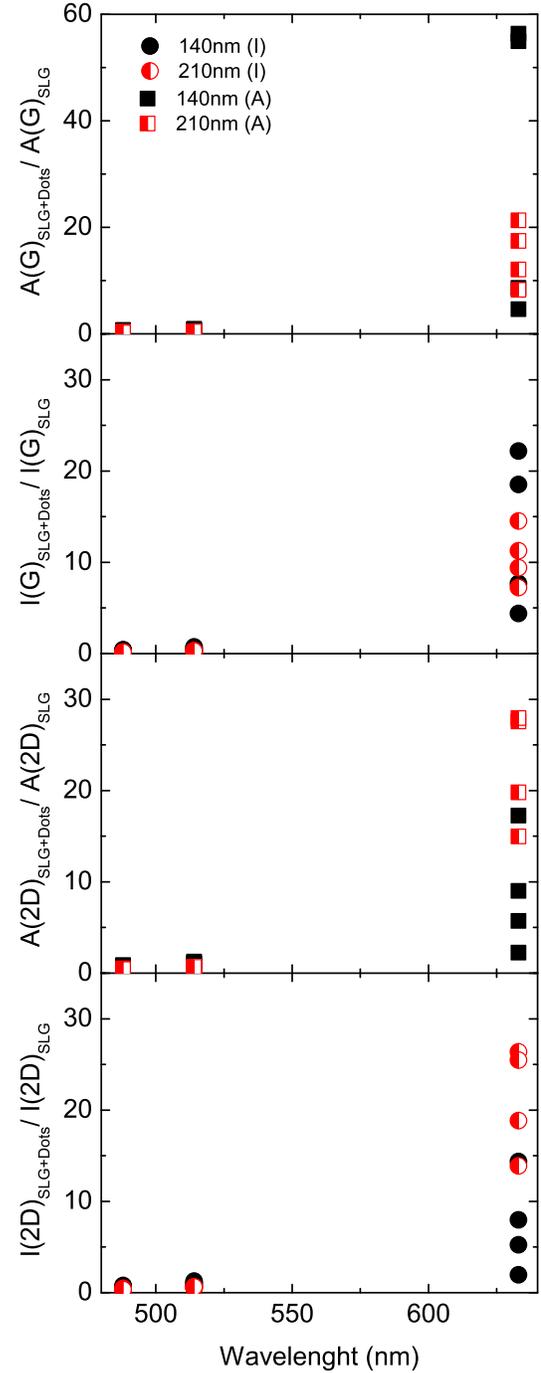}}
\caption{Ratio of height and areas of the G and 2D peaks measured on the patterned regions compared to those measured outside, as a function of the excitation energy}
\label{fig:fits}
\end{figure}

The Raman spectrum of graphene consists of a set of distinct peaks. The G and D appear around 1580 and 1350 cm$^{-1}$, respectively. The G peak corresponds to the $E_{2g}$ phonon at the Brillouin zone center ($\bf\Gamma$~point). The D peak is due to the breathing modes of six-atom rings and requires a defect for its activation\cite{tuinstra,Ferrari00,ThomsenPrl2000}. It comes from TO phonons around the \textbf{K} point\cite{tuinstra,Ferrari00}, is active by double resonance (DR)\cite{ThomsenPrl2000}, and is strongly dispersive with excitation energy due to a Kohn Anomaly at \textbf{K}\cite{PiscanecPRL}. DR can also happen as intra-valley process, i.~e. connecting two points belonging to the same cone around $\textbf{K}$ (or $\textbf{K}'$). This gives the so-called D'peak, which is at$\sim1620~cm^{-1}$ in defected graphite measured at 514nm. The 2D peak is the second order of the D peak. This is a single peak in single layer graphene (SLG), whereas it splits in four in bilayer graphene (BLG), reflecting the evolution of the band structure\cite{ACFRaman}. The 2D' peak is the second order of D'. Since 2D and 2D' originate from a process where momentum conservation is satisfied by two phonons with opposite wavevectors, no defects are required for their activation, and are thus always present.
Each Raman peak is characterized by its position, width, height, and area. The frequency-integrated area under each peak represents the probability of the whole process. It is more robust with respect to various perturbations of the phonon states than width and height\cite{baskoee}. Indeed, for an ideal case of dispersionless undamped phonons with frequency $\omega_\mathrm{ph}$, the shape of the $n$-phonon peak is a Dirac $\delta$ distribution $\propto\delta(\omega-n\omega_\mathrm{ph})$, with zero width, infinite height, but well-defined area. If the phonons decay (e.~g, into other phonons, due to anharmonicity, or into electron-hole pairs, due to electron-phonon coupling), the $\delta$ lineshape broadens into a Lorentzian, but the area is preserved, as the total number of phonon states cannot be changed by such perturbations. If phonons have a weak dispersion, states with different momenta contribute at slightly different frequencies. This may result in an overall shift and a non-trivial peak shape, but frequency integration across the peak means counting all phonon states, as in the dispersionless case. Thus, the peak area is preserved, as long as the Raman matrix element itself is not changed significantly by the perturbation. The latter holds when the perturbation (phonon broadening or dispersion) is smaller than the typical energy scale determining the matrix element. Converting this into a time scale using the uncertainty principle we have that, if the Raman process is faster than the phonon decay, the total number of photons emitted within a given peak (i.~e., integrated over frequency across the peak), is not affected by phonon decay, although their spectral distribution can be. Even if the graphene phonons giving rise to the D and D' peaks are dispersive\cite{PiscanecPRL}, their relative change with respect to the average phonon energy is at most a few \%, thus we are in the weakly dispersive case. The phonon decay in graphene is in the ps timescale, while the Raman process is faster, in the fs timescale\cite{Pisana,Bonini2007,LazzeriHP}. We thus consider both the area, A(2D)/A(G), and height, I(2D)/I(G), ratios\cite{baskoee}. Fig 4 plots them for the two dot sizes and as a function of excitation energy.

\begin{figure}
\includegraphics[width=80mm]{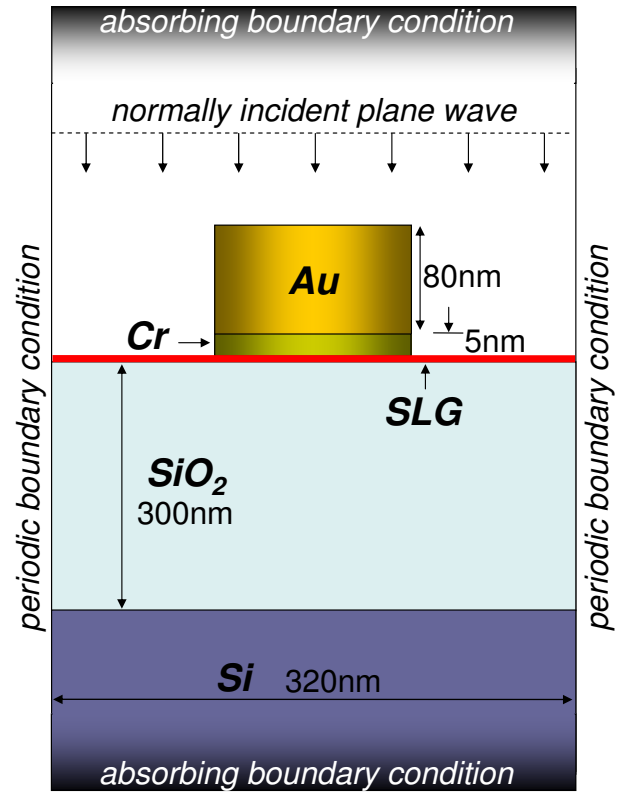}
\caption{Simulation box: 80nm/5nm Au/Cr nanodisks on SLG sitting on a SiO$_2$/Si. The lateral periodicity is 320nm in both directions. We consider nanodisk diameters of 210 and 140nm, corresponding to the large and small dots}
\label{fig:schematic}
\end{figure}

We model our experiment with the calculation box shown in Fig.~\ref{fig:schematic}. Starting from the bottom, this consists of a semi-infinite Si substrate, a 300nm SiO$_2$ and SLG of effective thickness 0.335nm. On SLG, we have a Au/Cr disk with thickness 80nm/5nm, with the diameter set to either 140 or 210nm, according to the experimental Au dot size. We time-integrate Maxwell's equations using the finite-difference time-domain method(FDTD)\cite{fdtd} as implemented in Refs\cite{ELnpJAP,ELcntACSNANO} (see Methods). For the absorbing boundary conditions in the vertical direction we use the perfectly-matched-layer method (PML)\cite{pml}, while in the lateral directions we use periodic boundary conditions simulating an infinite two-dimensional square array of Au/Cr nanodisks. We previously investigated the plasmonic resonances of similar nanoparticles (prepared in exactly the same conditions as in the present work)\cite{grigor, grigor2}. This allowed us to extract the Au and Cr optical constants as obtained in our evaporators\cite{kravets08}. We also recently measured the optical constants of graphene by spectroscopic ellipsometry\cite{kravets10}. The dispersive materials Au, Cr and graphene are described here by Drude-Lorentz models, each fitted to our experimental data. These are shown in Fig.~\ref{fig:epsilons} in Methods. Finally, for simplicity we use n=1.46 for SiO$_2$, n=4 for Si\cite{palik}.

We only consider normal incidence and emission, relevant for our Raman backscattering experiments. The incident field is a wide spectrum plane wave (i.e. a narrow Gaussian temporal profile) coming from the top. We monitor the electric fields $E_x(r,t)$ and $E_y(r,t)$ at each grid point on the graphene plane, and upon Fourier transform we get the tangential-field amplitude $E_{\parallel}(r,\omega)$ in frequency domain. This is enhanced compared to the incident field, due to substrate interference, and the Surface Plasmon Resonance (SPR) near field of the nanodisks.

Assuming the graphene absorption at a particular point to be proportional to the incident tangential field intensity at that point, the Raman emission from a particular point to be proportional to the corresponding Stokes-shifted intensity, and the emission from points underneath the nanodisks to be reabsorbed and lost, we may approximate the total Raman signal as\cite{moskovits85}:
\begin{equation}
\label{eq:approximation}
I_{SERS}\propto \int |E_{\parallel}(r,\omega)|^2 |E_{\parallel}(r,\omega_s)|^2 dS'
\end{equation}
where the integration is performed over the area not directly underneath the nanodisks, and $\omega_s = \omega - \delta\omega$ is the Stokes shifted frequency. $\delta\omega=2\pi c \nu$, with $\nu$ is the Raman shift (in cm$^{-1}$). The outcome of Eq.~\ref{eq:approximation} is normalized by the corresponding calculation for suspended graphene (where the integral is over all the area since there are no nondisks to cover the emission). We note that, given the large absorption of $\sim$2.3\% per graphene layer\cite{nairScience2008}, this approximation becomes questionable for local absorption enhancements greater than$\sim$43. To amend this we cut above 43. Emission saturation, on the other hand, cannot be reached because the Raman efficiency, even though larger than other materials due to the process being always resonant, is in absolute terms very small\cite{casiraghi2010}, and would require $\sim10^{11}$ enhancement to exceed 100\%.
\begin{figure}
\centerline{\includegraphics[width=85mm]{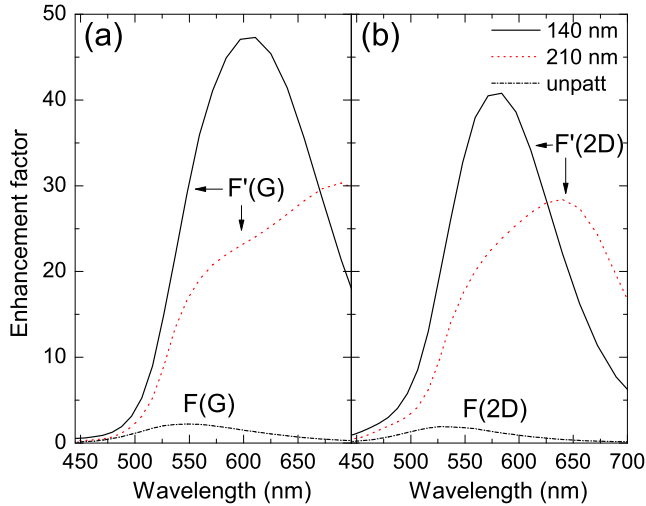}}
\caption{The total (patterned) enhancement factors for a) the G and b) 2D peaks. The dotted line is the corresponding interference (unpatterned) enhancement factor.}
\label{fig:enhancement1}
\end{figure}

In order to account for the graphene layer thickness and still keep the simulation reasonably fast, we employ an anisotropic grid with 0.335nm spacing in the vertical and 2nm in the lateral dimension. By calculations on smaller cells we verified that such a grid introduces small errors, typically less than 5\% and never exceeding 10\%. Considering our approximations up to this point (normal incidence and the approximation in Eq.~\ref{eq:approximation}), we find this to be well within the overall simulation errors.

The 300nm SiO$_2$/Si substrate interferometrically increases not only the visibility\cite{CasiraghiNL,BlakeAPL} but also the Raman signal of graphene\cite{wang2008,gao2009,berciaud}. Here we expect an additional enhancement due to the Au nanodisks SPR near field. In order to distinguish between the two effects (substrate interference and SPR), we separately calculate both cases for unpatterned SLG on SiO$_2$/Si and SLG patterned with the Au nanodisks, still on SiO$_2$/Si. We define the interference enhancement factor $F$ as the ratio of the Raman signal from unpatterned SLG on SiO$_2$/Si to that of suspended SLG, $F=I_{unpatt}/I_{susp}$, and the total enhancement factor $F'$ as the ratio of the Raman signal from patterned SLG on SiO$_2$/Si to that of suspended SLG: $F'=I_{patt}/I_{susp}\equiv I_{SERS}/I_0$. Fig.~\ref{fig:enhancement1} plots such factors for the G and 2D peaks. This gives a maximum interference-related enhancement $F$ at 550nm of$\sim$2.5 for the G peak and $\sim$2 for the 2D. However, this is very modest compared to the total enhancement $F'$ when the nanodisks are taken into account. The total enhancement reaches up to 50, and is maximum at different wavelengths depending on the nanodisk diameter. The shoulder at 550nm is likely related to interference, because (i) it is at the same frequency as the interference peak and (ii) it is stronger the further the plasmon peak.

\begin{figure}
\centerline{\includegraphics[width=85mm]{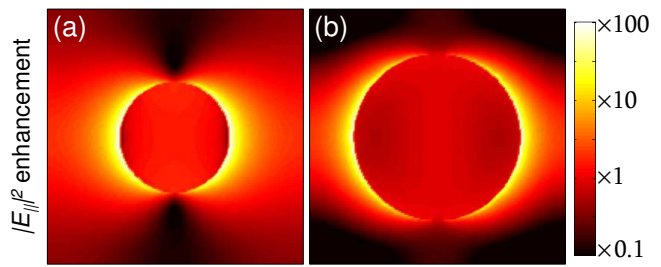}}
\caption{Tangential field intensity distribution at 633~nm for the a) 210~nm and b) 140~nm nanodisk diameters.}
\label{fig:intensitypatterns}
\end{figure}

There is a different enhancement for the two disk diameters, not only in peak value, but also in wavelength. This is a size effect: the nanodisks are comparable to the incident laser wavelength. Thus, retardation effects and higher order multipole terms become important, modifying the plasmon response as a function of disk size. Fig.~\ref{fig:intensitypatterns} plots the distribution of the tangential intensity enhancement in the graphene plane at 633nm, which is approximately proportional to the absorption enhancement at that wavelength. Different patterns appear for the two sizes, demonstrating the point made above.
\begin{figure}
\centerline{\includegraphics[width=85mm]{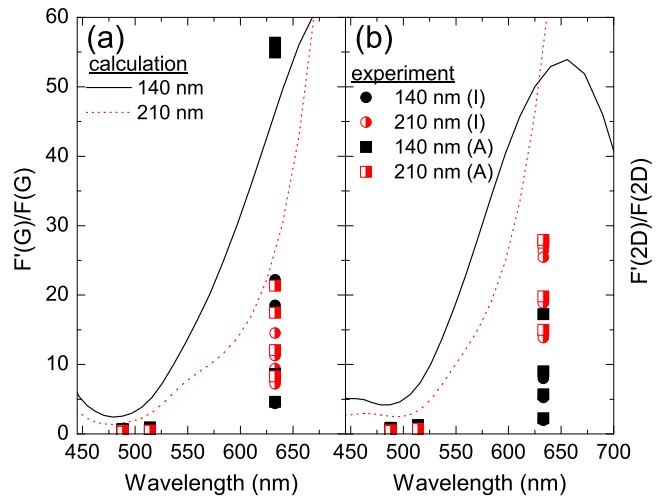}}
\caption{Normalized enhancement factors $F'/F$ for a) the G-band and b) the 2D-band. Squares and crosses are the experimental Raman intensities and areas.}
\label{fig:enhancement2}
\end{figure}

Fig.~\ref{fig:enhancement2} compares $F'/F$ to the corresponding experimental Raman intensity ratios. Considering all approximations made, and the fact that some of the measurements where on top of distorted parts of the nanodisk arrays, the agreement is good. We find low enhancement for 488 and 514nm, and large for 633nm. Also, we reproduce the higher enhancement for the G peak in the small disks at 633nm, and for the 2D peak in the large disks. The quantitative agreement for the 2D peak is not as good, however, this is expected given that its intensity significantly depends on electron-electron interactions, which could change in presence of gold\cite{baskoee}.

The agreement between experiment and simulation is encouraging. It also allows to get new physical insights into the SERS process of an extended 2d system, like graphene. The basic physics and detailed theory of the electromagnetic contribution to SERS on adsorbed molecules is well known\cite{moskovits85}: both absorption and emission are enhanced due to interaction with surface plasmons, with an expected overall dependence on the fourth power of the SPR-mediated field enhancement\cite{moskovits85,mccall80}. However, a detailed theory for 2d systems is lacking. Such formulation is challenging for our experiment: the Au particles do not have a regular spherical or ellipsoidal shape, they are large and so multipoles higher than dipole contribute, there is a thin Cr layer, we are on SiO$_2$/Si giving additional interference and enhancement effects. We thus consider the simplified case of regular-shaped small Au nanoparticles inside a uniform medium. This will provide all the new physics for SERS in graphene, or generally any 2d system. The final connection between experiment and theory still relies on simulations, the only versatile tool that can move across different scales and experimental embodiments.

\begin{figure}
\centerline{\includegraphics[width=95mm]{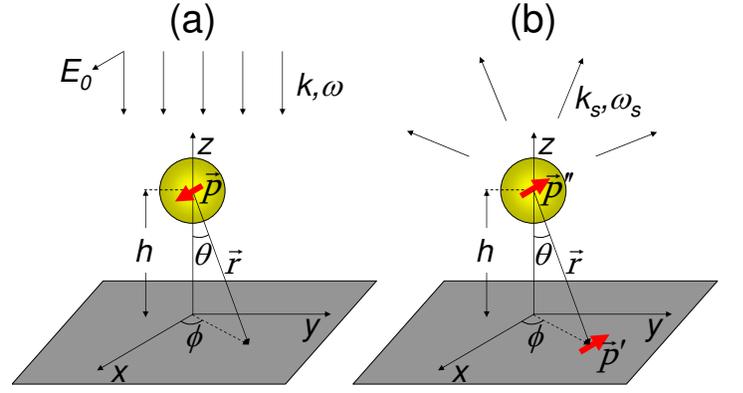}}
\caption{Scheme for SERS of a nanoparticle on graphene.}
\label{fig:schematic1}
\end{figure}

The generic theoretical system under study is depicted in Fig.~\ref{fig:schematic1}: at normal incidence, a plane wave of frequency $\omega$ excites a point dipole in the nanoparticle:
\begin{equation}
\label{eq:th1}
p \propto \alpha_{np}(\omega)
\end{equation}
where  the  polarizability $a_{np}$ is described by the Mie theory\cite{mie,bohren}. The poles of $\alpha_{np}$ define the optimal SERS frequencies. The re-radiated near-field from this dipole scales as $r^{-3}$, and is responsible for the enhanced absorption. This will excite a Raman dipole:
\begin{equation}
\label{eq:th3}
p^{\prime} \propto \alpha_R(\omega_s,\omega)r^{-3} \alpha_{np}(\omega)
\end{equation}
where $\alpha_R(\omega_s,\omega)$ is the Raman polarizability and $\omega_s$ the Stokes-shifted emission frequency. This dipole near-field will in turn excite a secondary dipole in the nanoparticle:
\begin{equation}
\label{eq:th4}
p^{\prime \prime} \propto \alpha_{np}(\omega_s)r^{-3}\alpha_R(\omega_s,\omega)r^{-3} \alpha_{np}(\omega)
\end{equation}
now at the emission frequency $\omega_s$. Thus, the additional surface-enhanced Raman signal is:
\begin{equation}
\label{eq:th5}
\Delta I_{SERS} \propto \omega_s^4 \int |p^{\prime \prime}|^2 dS
\end{equation}
where the integration is over the SLG area. Assuming a square array of nanoparticles with spacing $L$ and normalizing to the corresponding Raman signal in the absence of the nanoparticles, we get the SERS enhancement:
\begin{equation}
\label{eq:th6}
\frac{\Delta I_{SERS}}{I_0} \approx  \frac{1}{9} \sigma  Q(\omega)^2Q(\omega_s)^2 \left( \frac{a}{h}\right)^{10}
\end{equation}
where $h$ is the  separation between the nanoparticle center and the SLG plane, $\sigma=\pi a^2/L^2$ is the relative cross sectional area covered by the nanoparticles, and $Q(\omega)=\left| \alpha_{np}(\omega)\right| /4\pi a^3$ is the the Mie enhancement. In general, $Q$ depends on particle size and shape\cite{bohren} (for spherical particles $\alpha_{np}(\omega)$ is given by Eq.~\ref{eq:mie1} in Methods). In the limiting case of spherical particles with $a \ll \lambda$ the size dependence in the Mie enhancement is removed and $Q(\omega)\cong \left| [\epsilon(\omega)-1]/[\epsilon(\omega)+2]\right|$, where $\epsilon(\omega)$ is the nanoparticle's dielectric function.

This is our main theoretical result: the Raman enhancement scales with the metallic nanoparticle cross section, with the fourth power of the Mie enhancement, and inversely with the tenth power of the separation between the nanoparticle centre and graphene.
\begin{figure}
\centerline{\includegraphics[width=85mm]{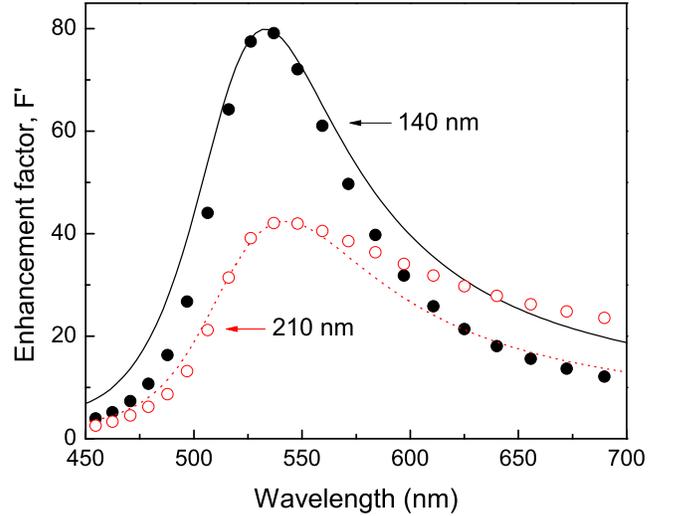}}
\caption{FDTD simulations of suspended nanodisks (symbols) compared with with Eq.~\ref{eq:th6} (lines) for G peak enhancement. The nanoparticle radius $a$ and elevation $h$ in Eq.~\ref{eq:th6} are adjusted to match the simulation peak enhancement. These are $2a$=118nm, $h$=32.2nm for the 140nm disks and $2a$=130nm, $h$=38nm for the 210nm disks. Note that the total enhancement factor $F'$ is just $F'=\Delta I_{SERS}/I_0 +1$.}
\label{fig:finalcomp}
\end{figure}

Since Eq.~\ref{eq:th6} does not take into account multiple reflections in the substrate, we compare it with FDTD simulations of Au nanodisks suspended in air, shown in  Fig.~\ref{fig:finalcomp} for the G peak (where there is a small blueshift due to the smaller effective refractive index below the nanoparticles without a substrate). In the absence of an analytical expression for the polarizability of a nanodisk, we use the polarizability of a sphere, shown in Eq.~\ref{eq:mie1} in Methods. To fit the simulation (symbols in Fig.~\ref{fig:finalcomp}) we adjust the radius $a$ (which determines the peak wavelength) and separation $h$ (which determines the peak value) in Eq.~\ref{eq:th6}. As a result, the 140nm nanodisk is best represented by a sphere of diameter $2a$=118nm elevated at $h$=32.2nm above the SLG plane, while for the 210nm nanodisk a sphere of diameter $2a$=130nm at $h$=38nm is needed (lines in Fig.~\ref{fig:finalcomp}). These numbers are close to the actual values for the nanodisks. The agreement is good, considering the difference between disks and spheres, confirming that our theory captures the essential physics of SERS in 2d.

Equation~\ref{eq:th6} is a valuable optimization/design tool. In particular, it identifies 3 steps to further improve SERS: 1) larger nanoparticle coverage $\sigma$, 2) larger Mie enhancement $Q$, 3) smaller nanoparticle-graphene separation $h$. The first is straightforward. The second is shape related, e.g., ellipsoids have a different $Q$, which, for certain orientations, is stronger and red-shifted compared to a sphere\cite{bohren}. This also influences the third step since, e.g., a flat oblate spheroid has a smaller distance $h$ between its center and graphene. Thus, SERS enhancements can be much larger for ellipsoids and disks. This can already be seen in the comparisons made in Fig.~\ref{fig:finalcomp}. There, in order to match the disk simulation results, the representative spheres had to be placed at an elevation $h<a$ (i.e. so that they "cut" into the SGL plane). If we, however, re-evaluate Eq.~\ref{eq:th6} for the same representative spheres at $h=a$ (i.e. so that they "rest" on the SLG plane), we lose more than two orders of magnitude enhancement. This is confirmed by detailed FDTD calculations.

Finally, we note that the analytical expressions derived in Methods can be extended to flakes of increasing number of layers, by vertical integration. Note, however, that if the number of layers is large reflection becomes important (it is ~1\% for 7 layers but exceeds 5\% above 18 layers) and has to be taken into account.
\section{conclusions}
We studied SERS in graphene patterned with a square array of Au nanodisks on SiO$_2$(300nm)/Si. Significant enhancements were measured for both G and 2D bands at 633nm. Similar results were obtained for both disk sizes. Large-scale FDTD simulations reproduce well the experiments. To elucidate the physics of SERS in 2d, we derived analytic expressions, and showed that taking into account the SPR near-fields only, a simple closed-form expression is found, where the Raman enhancement scales with the nanoparticle cross section, the fourth power of the Mie enhancement, and inversely with the tenth power of the separation between the nanoparticle centre and graphene. This points to thin nanodisks to achieve the highest SERS for 2d systems like graphene.
\section{acknowledgements}
ACF acknowledges funding from the Royal Society, the ERC grant NANOPOTS, and EPSRC grant EP/G042357/1. The authors acknowledge computing time at the Research Center for Scientific Simulations (RCSS) at the University of Ioannina.

\section{Methods}
Let us consider a generic configuration as in Fig.~\ref{fig:schematic1}, comprising a Au spherical nanoparticle of radius $a$ at a distance $h$ from SLG, normal plane wave incidence, with field amplitude $E_0$, frequency $\omega$ and polarization along $x$. The re-radiated fields from the nanoparticle are due to an induced electric dipole at its center:
\begin{equation}
\label{eq:dipole1}
\vec{p}=\epsilon_0 \alpha_{np}(\omega) \vec{E}\approx \epsilon_0 \alpha_{np}(\omega) E_0 \hat{\textbf{x}}
\end{equation}
where $E$ is the local field at the nanoparticle, modified from the incident one due to the presence of the graphene layer. The SLG normal incidence reflectance is almost zero\cite{nairScience2008} and, due to continuity, the local field can be taken approximately the same as the incident one. This is further corroborated from calculations of the depolarization matrix of single wall nanotubes, where for polarization along the nanotube axis, no depolarization is found\cite{cohen1995,cancado2009}. The nanoparticle polarizability $a_{np}$ is described for a sphere in the Mie theory\cite{mie,bohren}:
\begin{equation}
\label{eq:mie1}
\alpha_{np}(\omega)=\frac{6\pi i }{k^3} \cdot
\frac{\tilde{n}\psi_1(\tilde{n} ka) \psi_1^{\prime}(ka) - \psi_1(ka) \psi_1^{\prime}(\tilde{n}ka)}
{\tilde{n}\psi_1(\tilde{n} ka) \xi_1^{\prime}(ka) - \xi_1(ka) \psi_1^{\prime}(\tilde{n}ka)}
\end{equation}
where $\psi_1$ and $\xi_1$ are the Riccati-Bessel functions, $k=\omega/c$ and $\tilde{n}\equiv \tilde{n}(\omega)$ is the nanoparticle complex index of refraction. In general, the refractive index of metallic nanoparticles differs from the bulk value due the reduced free electron relaxation caused by electron surface scattering\cite{kreibig,ELnpJAP}. We have relatively large particles, where this correction is negligible. Thus we use the bulk Au refractive index. If the nanoparticle is much smaller than the wavelength, Eq.~\ref{eq:mie1} simplifies\cite{bohren}:
\begin{equation}
\label{eq:mie2}
\alpha_{np}(\omega) \approx 4\pi a^3 \frac{\epsilon(\omega)-1}{\epsilon(\omega)+2}
\end{equation}
with $\epsilon(\omega)=\tilde{n}^2(\omega)$ the nanoparticle dielectric function.

The total field at at position $\vec{r}$ on SLG is\cite{jackson}:
\begin{eqnarray}
\label{eq:totalfield}
&&\vec{E}_t(\vec{r},\omega) \approx E_0(\vec{r},\omega) \hat{\textbf{x}}+\nonumber\\
&&\frac{\mathrm{e}^{ikr}}{4\pi\epsilon_0} \left[\frac{k^2}{r} \hat{\textbf{r}}\times(\vec{p}\times \hat{\textbf{r}}) + \left( \frac{1}{r^3}-\frac{ik}{r^2} \right)[3\hat{\textbf{r}}(\hat{\textbf{r}}\cdot\vec{p})-\vec{p}]\right]\nonumber\\
\end{eqnarray}
The first term in the square bracket is the radiation-field, while the second is the near-field. They scale with distance and wavelength as $(\lambda^2 r)^{-1}$ and $r^{-3}$, $(\lambda r^2)^{-1}$ respectively. Assuming the nanoparticle distance from SLG to be much smaller than the wavelength, $r \ll \lambda$, and the resonant near-field much larger than the incident field $E_0$, then the near-field $r^{-3}$ has the dominant contribution. Its transverse component drives the SLG enhanced absorption. Simultaneously, the Raman emitted field also interacts with the Au particle SPR, further enhancing the total Raman emission. To fully explore both processes, we evaluate two quantities: 1) total absorption enhancement; 2) total Raman enhancement.

\subsection{Absorption enhancement}
The additional absorption in graphene is due to the enhanced near field. Absorption is defined as current$\times$field\cite{jackson}, thus approximated as:
\begin{equation}
\label{eq:absorption1}
\Delta A= G_0 \int |\vec{E}^{nf}_{t_{\parallel}}|^2 dS
\end{equation}
where $G_0=e^2/4\hbar$ is graphene's dynamical (optical) sheet conductance\cite{nairScience2008,kuzmenko08,ando02,gusynin06,falkovsky07}. Combining Eqs.~\ref{eq:dipole1},\ref{eq:totalfield}, the near-field transverse component of the driving field is:
\begin{equation}
\label{eq:field1}
|\vec{E}^{nf}_{t_{\parallel}}(\omega)| = E_0 Q(\omega)a^3r^{-3}|[3\hat{\textbf{r}}(\hat{\textbf{r}}\cdot\hat{\textbf{x}})-\hat{\textbf{x}}]_{\parallel}|
\end{equation}
with $Q(\omega)=\left| \alpha_{np}(\omega) \right| /4\pi a^3$. In the above we ignore cross terms between incident and re-radiated fields. This is justified when strong near-field enhancements are expected (as it should be for SERS), but less so when the near fields are of the same order as the incident field.

From Fig.~\ref{fig:schematic1}, $x=r\sin{\theta}\cos{\phi}$, $y=r\sin{\theta}\sin{\phi}$ and $z=r\cos{\theta}$, while $r=h/\cos{\theta}$. The integration surface element is $\rho d\rho d\phi = h^2 \sin{\theta}\cos^{-3}{\theta}d\phi d\theta$, where $\rho=h\tan{\theta}$ is $\vec{r}$'s projection on the plane. Eq.~\ref{eq:absorption1} then becomes:
\begin{equation}
\label{eq:absorption2}
\Delta A= \frac{ G_0 E_0^2Q^2 a^6}{h^4} \int_{0}^{\frac{\pi}{2}} \int_{0}^{2\pi}
\cos^3{\theta} \sin{\theta} f(\theta,\phi)d\phi d\theta
\end{equation}
with $f(\theta,\phi)=9\sin^4{\theta}\cos^2{\phi}-6\sin^2{\theta}\cos^2{\phi}+1$. Then:
\begin{equation}
\label{eq:absorption3}
\Delta A= \frac{ 3 \pi G_0 E_0^2Q^2(\omega) a^6}{8 h^4}
\end{equation}
For nanoparticles arranged on a square lattice with spacing $L$, the absorption enhancement becomes:
\begin{equation}
\label{eq:absorptionenh}
\frac{\Delta A}{A_0}=\frac{3\pi Q^2(\omega) a^6}{8h^4 L^2} = \frac{3}{8}\sigma  Q^2(\omega) \left(\frac{a}{h}\right)^4
\end{equation}
where $\sigma=\pi a^2/L^2$ is the nanoparticle relative cross section.
For spheres directly placed on SLG, i.e. $h=a$, Eq.~\ref{eq:absorptionenh} simplifies to:
\begin{equation}
\label{eq:absorptionenh2}
\frac{\Delta A}{A_0}= \frac{3}{8} \sigma Q^2(\omega)
\end{equation}
We remind here that in the $a\ll \lambda$ limit, the Mie enhancement is $Q(\omega)=\left|[\epsilon(\omega) -1]/[\epsilon(\omega) + 2]\right|$.
\subsection{Raman enhancement}
Going back to Eq.~\ref{eq:field1}, $\vec{E}^{nf}_{t_{\parallel}}(\omega)$ will excite a dipole field on SLG at the Raman frequency $\omega_s$:
\begin{equation}
\label{eq:ramandipole}
\vec{p}^{\prime}=\alpha_R(\omega_s,\omega) |\vec{E}^{nf}_{t_{\parallel}}| \hat{\textbf{p}}^{\prime}
\end{equation}
The polarization of the Raman dipole is not necessarily the same as that of the driving field. Thus, for generality we assume this to be randomly polarized on the SLG plane. In this case it suffices to take the average of two dipoles, one polarized along $\hat{\textbf{x}}$, and another along $\hat{\textbf{y}}$. The one along $\hat{\textbf{x}}$ will emit as the dipole term of Eq.~\ref{eq:totalfield}.  We are again interested in the dominant near-field term that decays as $r^{-3}$. This will get coupled to the nanoparticle and thus SPR enhanced. That is, it will excite a secondary dipole at the nanoparticle:
\begin{equation}
\label{eq:field2}
p^{\prime \prime} = Q(\omega_s) \alpha_{R}(\omega_s,\omega) Q(\omega)  E_0 a^6 r^{-6} |[3\hat{\textbf{r}}(\hat{\textbf{r}}\cdot\hat{\textbf{x}})-\hat{\textbf{x}}]_{\parallel}|^2
\end{equation}
where we again consider the projection of the dipole given the backscattering geometry considered here. The radiated flux can be taken as the additional surface-enhanced Raman signal:
\begin{equation}
\label{eq:raman}
\Delta I_{SERS} = \frac{ck_s^4}{24 \pi \epsilon_0} \int p^{\prime \prime 2} dS
\end{equation}
where $k_s=\omega_s/c$ and we multiply by a factor 1/2 since we only consider the upper half flux. Using the angular relationships for $x,y,z$ and $dS$ we get:
\begin{eqnarray}
\label{eq:raman1}
&&\Delta I_{SERS} = \frac{ck_s^4E^2_0  Q^2(\omega) Q^2(\omega_s)a^{12}}{24 \pi \epsilon_0 h^{10}} |  \alpha_{R}(\omega_s,\omega)   |^2 \times\nonumber\\
&&\int_{0}^{\frac{\pi}{2}} \int_{0}^{2\pi}
\cos^9{\theta}\sin{\theta} f_x(\theta,\phi) d\phi d\theta
\end{eqnarray}
where $f_x(\theta,\phi)=81\sin^8{\theta}\cos^4{\phi}-108\sin^6{\theta}\cos^4{\phi}+18\sin^4{\theta}\cos^2{\phi}(1+2\cos^2{\phi})-
12\sin^2{\theta}\cos^2{\phi}+1$. The angular integration yields $ 33\pi/280$. For the calculation with the Raman dipole along $\hat{\textbf{y}}$, the angular part of Eq.~\ref{eq:field2} becomes $|[3\hat{\textbf{r}}(\hat{\textbf{r}}\cdot\hat{\textbf{x}})-\hat{\textbf{x}}]_{\parallel}|
|[3\hat{\textbf{r}}(\hat{\textbf{r}}\cdot\hat{\textbf{y}})-\hat{\textbf{y}}]_{\parallel}| $ with a slightly different $f_y(\theta,\phi)$, but a similar angular integration value of $27\pi/280$. The average angular contribution is $3\pi/28 \approx \pi/9$. Thus the additional Raman signal can thus be written as:
\begin{equation}
\label{eq:ramanfinal}
\Delta I_{SERS}  \approx \frac{\pi}{9}\frac{c k_s^4 E^2_0 Q^2(\omega) Q^2(\omega_s) a^{12}}{24 \pi \epsilon_0 h^{10}}  | \alpha_{R}(\omega_s,\omega)   |^2
\end{equation}

To evaluate the enhancement factor, we normalize to the expected signal $I_0$ in the absence of the nanoparticles. Considering a square unit cell of side equal to the nanoparticle spacing $L$, we get:
\begin{equation}
\label{eq:ramanbare}
I_0\approx L^2 \frac{ck_s^4E^2_0}{24 \pi \epsilon_0}| \alpha_{R}(\omega_s,\omega)   |^2
\end{equation}
The Raman enhancement factor is
\begin{equation}
\label{eq:ramanenhancement}
 \frac{\Delta I_{SERS}}{I_0} \approx  \frac{1}{9}\sigma  Q^2(\omega) Q^2(\omega_s) \left( \frac{a}{h} \right)^{10}
\end{equation}
\subsection{The Finite-Difference Time-Domain Method}
\begin{figure}
\centerline{\includegraphics[width=90mm]{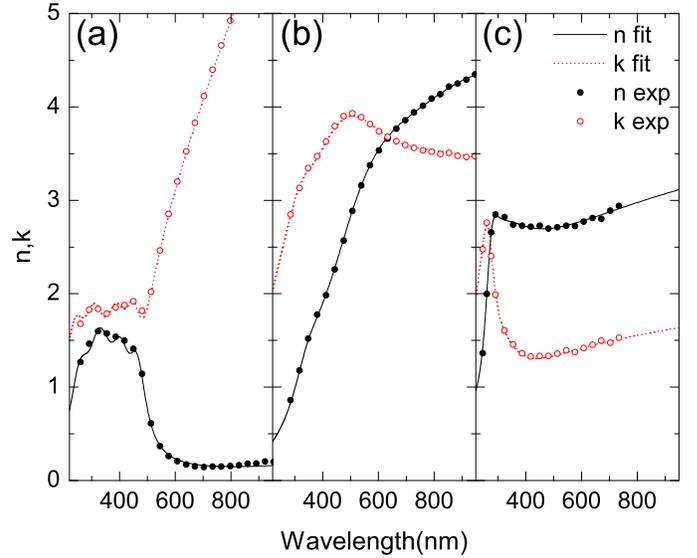}}
\caption{The refractive indices used in the calculations for a) Au, b) Cr and c) SLG. Symbols are for the corresponding experimental data\cite{kravets08,kravets10}.}
\label{fig:epsilons}
\end{figure}

In the FDTD method, Maxwell's equations are time-integrated on a computational grid:
\begin{eqnarray}
\nabla \times \textbf{E}&=&-\mu \partial_t\textbf{H} \\
\nabla \times \textbf{H}&=& \epsilon_0 \epsilon_{\infty} \partial_t\textbf{E}+\partial_t\textbf{P}_0 +\sum_{j=1}^{N}\partial_t \textbf{P}_j
\end{eqnarray}
where material polarization is taken into account through the polarizabilities \textbf{P}:
\begin{eqnarray}
\partial_t^2 \textbf{P}_0+\gamma \partial_t\textbf{P}_0&=&\omega_p^2\epsilon_0\textbf{E} \\
\partial_t^2\textbf{P}_j+\Gamma_j \partial_t \textbf{P}_j+\Omega_j^2\textbf{P}_j&=&\Delta \epsilon_j\Omega_j^2\epsilon_0\textbf{E}
\end{eqnarray}
This gives a Drude-Lorentz model for the dielectric function\cite{wooten}:
\begin{equation}
\epsilon(\omega)=\epsilon_{\infty}-\frac{\omega_p^2}{\omega^2+i\omega\gamma}
+\sum_{j=1}^N\frac{\Delta\epsilon_j\Omega_j^2}{\Omega_j^2-\omega^2-i\omega\Gamma_j}
\end{equation}
where the first term is the Drude free-electron contribution and the second contains Lorentz oscillators corresponding to interband transitions. $\omega_p$ and $1/\gamma$ are the free electron plasma frequency and relaxation time, $\Omega_j$, $\Delta\epsilon_j$, $\Gamma_m$ are transition frequency, oscillator strength and decay rate for the Lorentz terms. To accurately reproduce the experimental dielectric functions (Au and Cr from Ref.\cite{kravets08}, SLG from Ref.\cite{kravets10}) we treat these as fit parameters.  For Au we use $N$=4, and $\epsilon_\infty$=3.454, $\Delta\epsilon_j$=(0.376, 0.63, 1.208, 1.124), $\hbar\omega_p$=8.73eV, $\hbar\gamma$=0.046eV, $\hbar\Omega_j$=(2.72, 3.13, 3.88, 4.95)eV, and $\hbar\Gamma_j$=(0.39, 0.655, 1.16, 1.67)eV.
For Cr we use $N$=3, and $\epsilon_\infty$=1, $\Delta\epsilon_j$=(9.54, 15.5, 1.1), $\hbar\omega_p$=5.51eV, $\hbar\gamma$=0.731eV, $\hbar\Omega_j$=(1.43, 2.36, 3.64)eV, and $\hbar\Gamma_j$=(1.19, 1.94, 1.41)eV.
Finally, for SLG we use $N$=3, and $\epsilon_\infty$=1.964, $\Delta\epsilon_j$=(6.99, 1.69, 1.53), $\hbar\omega_p$=6.02eV, $\hbar\gamma$=4.52eV, $\hbar\Omega_j$=(3.14, 4.03, 4.59)eV, and $\hbar\Gamma_j$=(7.99, 2.01, 0.88)eV.
Fig.~\ref{fig:epsilons} plots our model dielectric functions along with the experimental ones, showing an excellent agreement.

\end{document}